\input epsf

\documentclass[draft]{aa}

\begin{document}

\thesaurus{20(08.05.2, 08.06.3, 08.18.1)} 
 
\title
{A Representative Sample of Be Stars I:  Sample Selection, Spectral Classification and Rotational Velocities}

\author{I.~A.~Steele\inst{1} 
\and I.~Negueruela\inst{1,2} 
\and J.~S.~Clark\inst{3}
}                   
                                                            
\institute{Astrophysics Research Institute, Liverpool John Moores 
University, Twelve Quays House, Egerton Wharf, Birkenhead, L41 1LD, UK
\and SAX SDC, Italian Space Agency, c/o Telespazio, via Corcolle 19, I00131,
Rome, Italy
\and Astronomy Centre, CPES, University of Sussex, Brighton, BN1 9QH, UK
}

\offprints{I. A. Steele}
\mail{ias@astro.livjm.ac.uk}

\date{Received 7 September 1998 / Accepted 12 March 1999   }

\titlerunning{A Representative Sample of Be stars I}
\authorrunning{Steele, Negueruela \& Clark}
\maketitle 

\begin{abstract}

We present a sample of 58 Be stars containing objects
of spectral types O9 to B8.5 and luminosity classes III to V.  
We have obtained 3670 -- 5070 {\AA} spectra of the sample which are
used to derive spectral types and rotational velocities.  We discuss the
distribution of spectral types and rotational velocities obtained
and conclude that there are no significant selection effects in
our sample.

\end{abstract}


\section{Introduction}

Fundamental questions still outstanding with regard
to Be stars are the radial dependence of the density, temperature
and velocity structure of the circumstellar disk surrounding the star.  In
addition the dependence of these parameters on the 
effective temperature (spectral type),
evolutionary status (luminosity class), and rotational velocity 
of the underlying B star is unknown.  
In an attempt to answer these questions we have devised a multiwavelength 
approach which combines optical and infrared spectroscopy of a carefully 
selected sample of Be stars containing both giants and dwarfs of
spectral types from O9 to B8.5.   Our dataset comprises observations of
58 Be stars in 5 wavelength regions and is summarised in Table 1.

\begin{table*}
\caption{Summary of observations of the sample}
\begin{tabular}{ccll}

Wavelength Range ({\AA}) & Resolution (km/s) & Date Obtained & Emission Lines Seen \\
\hline
3670 -- 5070 & $\sim 55$  & 1998 Aug  & H$\beta$ - H limit, He{\sc i}, Mg {\sc ii}, Fe {\sc ii} \\
5850 -- 7150 & $\sim 35$  &  1998 Aug  & H$\alpha$, He {\sc i}  \\
7800 - 9200  & $\sim 30$  & 1998 Aug  & Paschen 6 -- Paschen limit, He {\sc i} \\
15300 - 16900 &$\sim 100$  &1996 Jun + Oct & Brackett 11 to 18  \\
20550 - 22100 & $\sim 70$  &1996 Jun + Oct & Mg {\sc ii}, He {\sc i}, Br $\gamma$, Fe {\sc ii}, Na {\sc i} \\ 
\end{tabular}
\end{table*}

This paper presents the details of the sample selection and
an analysis of the classification (absorption line) features in the
3670 -- 5070 {\AA} spectra.  We defer discussion of the emission line
features in these spectra to subsequent papers, where we will consider
them in conjunction with the emission line spectra at other wavelengths.  

\section{Sample Selection}

A Be star is defined as a B-type non-supergiant star that shows, or has
shown in the past, emission lines.  Jaschek \& Egret (1982) provide
a list of such objects, and it is from that list our sample is drawn.
We note here that as the Be phenomenon is time variable
we expect several of our objects not to show emission lines at present.

Our sample is termed a ``representative'' sample, in that it
was selected in an attempt to contain several objects that were 
typical of each spectral and luminosity class for which the
Be phenomenon occurs.  It therefore does {\em not}
reflect the spectral and luminosity class space distribution of Be stars,
but only the attempts to define the
average properties of each subclass in temperature and
luminosity.  Our selection was made according to the following criteria:

\begin{enumerate}
\item an equal distribution of spectral types from B0 to B9 
using the spectral types listed by Jaschek \& Egret (\cite{je82})
\item an equal distribution between dwarf and giant luminosity 
classes listed by Jaschek \& Egret (\cite{je82})
\item no evidence of spectroscopic binarity in the literature
\end{enumerate}

The following additional constraints were imposed by the instrument 
sensitivity and
time allocation:

\begin{enumerate}
\item Right Ascensions in the range 17$^h$ to 6$^h$
\item Declinations in the range to +59$^{\circ}>\delta> -27^{\circ}$
\item $B$ magnitude brighter than $\sim 11$
\end{enumerate}
  
These criteria were designed in an attempt to
create a sample containing roughly 3 objects per spectral type 
per luminosity class.  We note here that no selection criteria was
applied for $v \sin i$, as the distribution of $v \sin i$ with
spectral type and luminosity class was one of the phenomena
we wished to investigate.
The achieved spectral and luminosity class distribution is somewhat different
to the 3 per spectral type per luminosity class,
with many more dwarfs than giants in the sample.  This is not
surprising given that Jaschek \& Egret (1982) is simply a compilation
of values from the literature, with most of the spectral types from
low resolution photographic spectra taken in the 1950's.  The
achieved distribution 
is discussed in 
Sections 4 and 5,
along with a discussion of possible biases in the sample.

\begin{table*}
\caption{Data on the sample from the SIMBAD database (Coordinates J2000)}
\begin{tabular}{lllllll}
OBJECT & RA & DEC & $B$ & $V$ & Alias & Historical Spectral Type(s) \\
\hline
CD -28 14778 & 18 37 40.0  & -27 59 07   & 9.06 & 8.95 & HD 171757 & B2nne, B3IIIe  \\
CD -27 11872 & 17 44 45.7  & -27 13 44   & 9.13 & 8.69 & V3892 Sgr, HD 161103 &  Be, B2:IVpe, B2III-IVpe \\
CD -27 16010 & 22 40 39.2  & -27 02 37   & 4.06 & 4.20 & $\epsilon$ PsA, HR 8628, HD 214748 & B8V, B7,  B8Ve   \\
CD -25 12642 & 18 03 44.0  & -25 18 54   & 9.29 & 9.00 & HD 164741 & B2Ib/II, B1III \\
CD -22 13183 & 18 39 30.1  & -21 57 56   & 8.08 & 7.90 &  HD 172158 & B8II  \\
BD -20 05381 & 19 03 33.1  & -20 07 42   & 7.76 & 7.80 &  HD 177015 & B4Vn, B5III \\
BD -19 05036 & 18 31 24.1  & -19 09 31   & 8.26 & 7.92 & V3508 Sgr, HD 170682 & B5II/III,  B7III \\
BD -12 05132 & 18 39 39.7  & -11 52 43   & 10.13 & 9.48 & HD 172252 & B0Ve, B2Vnpe   \\
BD -02 05328 & 20 39 13.1  & -02 24 46   & 6.12 & 6.22   & HD 196712 & B7IIIne B8Ve  \\
BD -01 03834 & 19 49 33.4  & -01 06 03   & 8.25 & 8.14   & HD 187350 & B1Vne   \\
BD -00 03543 & 18 44 55.8  & -00 22 23   & 6.87 & 6.88  & HD 173371 & B9III \\
BD +02 03815 & 19 12 03.2  & +02 37 22   & 7.03 & 6.92 & HD 179343 & B9   \\
BD +05 03704 & 18 21 28.4  & +05 26 09   & 6.09 & 6.13  & HD 168797 & B3Ve, B2Ve   \\
BD +17 04087 & 19 46 57.8  & +18 14 56   & 10.6 & 10.6 & HD 350559 & B7IIIe    \\
BD +19 00578 & 03 42 18.9  & +19 42 02   & 5.68 & 5.69 & 13 Tau, HR 1126, HD 23016 & B9Vn, B8Ve \\
BD +20 04449 & 20 09 39.6  & +21 04 44   & 8.3  & 8.3  & HD 191531 & B0.5 III-V  \\
BD +21 04695 & 22 07 50.4  & +21 42 14   & 5.68 & 5.78 & 25 Peg, HD 210129 & B7Vn,   B6V, B8V \\
BD +23 01148 & 06 01 05.8  & +23 20 20   & 7.97 & 7.36 & HD 250289 & B2IIIe  \\
BD +25 04083 & 20 04 00.7  & +26 16 17   & 9.57 & 8.94 & HD 339483 & B1III \\
BD +27 00797 & 05 34 36.6  & +27 35 34   & 10.31 & 9.86 & HD 244894 & B1III-IVpe  \\
BD +27 00850 & 05 44 27.7  & +27 13 48   & 9.54 & 9.38 & HD 246878 &  B0.5Vpe  \\
BD +27 03411 & 19 30 45.3  & +27 57 55   & 5.01 & 5.15 & $\beta$2 Cyg, HR 7418, HD 183914 & B8V,  B7V   \\
BD +28 03598 & 20 03 11.7  & +28 42 27   & 10.45 & 9.43 & HD 333452 & B0III:np \\
BD +29 03842 & 20 00 33.5  & +30 22 52   & 10.60 & 10.11 & HD 33226  & B1Ve \\
BD +29 04453 & 21 35 44.4  & +29 44 44   & 8.15 & 8.10 & HD 205618 & B2Vne   \\
BD +30 03227 & 18 33 23.0  & +30 53 32   & 6.47 & 6.58 & HR 6971, HD 171406 & B4V, B3Vn \\
BD +31 04018 & 20 16 48.2  & +32 22 48   & 7.25 & 7.16 & V2113 Cyg, HD 193009 &  B1V:nnpe   \\
BD +36 03946 & 20 13 50.2  & +36 37 23   & 9.0  & 9.2 & HD 228438 & B0.5III, B0IV  \\  
BD +37 00675 & 03 00 11.7  & +38 07 55   & 6.05 & 6.16 & HR 894, HD 18552 & B8Vn , B8Ve, B9V  \\
BD +37 03856 & 20 16 08.4  & +37 33 23   & 10.58 & 10.20 & HD 228650 & B1V  \\
BD +40 01213 & 05 13 13.3  & +40 11 37   & 7.38 & 7.34 & HD 33604 &  B2Vpe   \\
BD +42 01376 & 05 42 19.9  & +43 03 35   & 7.27 & 7.28 & V434 Aur, HD 37657 & B3Vne  \\
BD +42 04538 & 22 55 47.0  & +43 33 34   & 8.06 & 8.02 & HD 216581 & B3Vn      \\
BD +43 01048 & 04 45 51.4  & +43 59 40   & 9.81 & 9.53 & HD 276738 &  B7V  \\
BD +45 00933 & 04 25 49.9  & +46 14 02   & 8.38 & 8.06 & HD 27846 & B1.5V    \\
BD +45 03879 & 22 19 00.2  & +45 48 08   & 8.48 & 8.48 & HD 211835 & B3Ve, B2n   \\
BD +46 00275 & 01 09 30.1  & +47 14 31   & 4.18 & 4.25 & $\phi$ And, HR 335, HD 6811 & B7III, B7IV , B7V , B8III  \\
BD +47 00183 & 00 44 43.4  & +48 17 04   & 4.47 & 4.50 & 22 Cas, HR 193, HD 4180 & B5III, B2V, B3IV, B2Ve  \\
BD +47 00857 & 03 36 29.2  & +48 11 35   & 4.17 & 4.23 & $\psi$ Per, HR 1087, HD 22192 & B5Ve, B5ne, B5IIIe, B5V \\
BD +47 00939 & 04 08 39.5  & +47 42 47   & 4.01 & 4.04 & 48 Per, HR 1273, HD 25940 & B3Ve, B3p , B3Vpe \\
BD +47 03985 & 22 57 04.4  & +48 41 03   & 5.33 & 5.42 & EW Lac, HR 8731, HD 217050 & B3IVpe , B2pe , B5ne  \\
BD +49 00614 & 02 16 36.0  & +49 49 12   & 7.59 & 7.57 & HD 13867 & B5Ve,  B8e  \\
BD +50 00825 & 03 48 18.2  & +50 44 12   & 6.20 & 6.15 & HR 1160, HD 23552 & B8Vn, B7V  \\
BD +50 03430 & 21 46 02.7  & +50 40 27   & 6.96 & 7.02& HD 207232 & B9  \\
BD +51 03091 & 21 34 27.4  & +51 41 54   & 6.17 & 6.19 & HR 8259, HD 20551 & B9III  \\
BD +53 02599 & 21 19 44.8  & +53 57 06   & 8.34 & 8.08 & HD 203356 & B9 \\
BD +55 00552 & 02 15 02.5  & +55 47 36   & 8.25 & 7.90  & HD 13669 & B3IV-V, B2Vne, B2V  \\
BD +55 00605 & 02 23 35.4  & +56 34 28   & 9.61 & 9.34 & V361 Per, HD 14605 &  B0.5Vpe, B1.5IIIe  \\
BD +55 02411 & 20 29 27.0  & +56 04 06   & 5.87 & 5.89 & HD 195554 & B9Vn  \\
BD +56 00473 & 02 16 57.6  & +57 07 49   & 9.33 & 9.08 & V356 Per & B0.5IIIn, B1III, B1II, B3e  \\
BD +56 00478 & 02 17 08.1  & +56 46 11   & 8.70 & 8.51 & V358 Per, HD 13890 & B1III, B1IIIpe, B3e \\
BD +56 00484 & 02 17 44.6  & +56 54 00   & 9.94 & 9.62 & V502 Per &  B1IIIe, B0Vne, B0ne  \\
BD +56 00493 & 02 18 18.0  & +56 51 03   & 9.81 & 9.62 & - &  B1Vpe \\
BD +56 00511 & 02 18 47.9  & +57 04 02   & 9.49 & 9.11 & - &  B3III  \\
BD +56 00573 & 02 22 06.4  & +57 05 25   & 10.06 & 9.66 & - &  B2III - IVe  \\
BD +57 00681 & 03 02 37.8  & +57 36 46   & 9.34  & 8.66 & HD 237056 & B0.5Bpe, 08ne  \\             
BD +58 00554 & 03 04 30.5  & +59 26 47   & 9.49 & 9.16 & HD 237060 & B9V \\    
BD +58 02320 & 21 44 33.9  & +59 03 26   & 9.77 & 9.51 & HD 239758 &  B2III:nn, B2Vn(e)  \\
\end{tabular}
\end{table*}

There are a total of 58 objects in our sample.
In Table 2 we list the sample by BD number, along with other aliases
for the brighter/well studied objects. 
Also listed in Table 2 are J2000 coordinates and $B$ and $V$ magnitudes
for the sample from the SIMBAD database, and the historical spectral
types assigned to the objects.  

Note that there is a large variation in the historical 
spectral types, with many objects having a spectral class uncertain by
2 sub-types and a  similarly uncertain luminosity class.  We assume this
is due to the quality of the spectra used.  Although Be stars show great 
time variability in
their emission line spectra, we are aware of no evidence in the modern literature to indicate variability of the underlying B star absorption spectrum.  
Since the overall aim of this programme is to understand the relationship
between the parameters of underlying stars and the circumstellar disks in
these systems it is therefore necessary to reclassify all of the stars
in our sample using modern CCD spectra.  This need to reclassify the
sample is the motivation for the observations presented here.

\section{Observations}

The classification spectra were obtained using the IDS spectrograph of
the Isaac Newton Telescope, La Palma on the night of 1998 August 2.
The R1200B grating was employed with a slit width of 1.15 arcsec and the
EEV12 CCD.  This gives a dispersion of $\sim 0.5$ {\AA}/pixel.
A central wavelength was chosen of 4300 {\AA}, giving a
a wavelength range of 3670 - 5070 {\AA}.  Measurements of
interstellar features in the spectra give a full width half maximum
corresponding to a velocity resolution of $\sim 55$ km/s.
In addition to our
Be star sample we obtained spectra of
30 MK standards in the range O9 to B9.5, mainly of luminosity classes
III and V (Table 3). 

\begin{table}
\caption{Spectral Standards observed ($^*$ = see text)}
\begin{tabular}{lll}
HD Number & Alias & Spectral Type \\
\hline
 HD 214680  & 10 Lac  &  O9V\\
 HD 34078  & AE Aur  &  O9.5V\\
 HD 209975  & 19 Cep  &  O9.5Ib \\
 HD 149438  & $\tau$ Sco   &  B0.2V\\
 HD 22951  & 40 Per  &  B0.5V\\
 HD 218376  & 1 Cas  &  B0.5III\\
 HD 144470  & $\omega^{1}$ Sco   &  B1V\\
 HD 23180  & $o$ Per  &  B1III\\
 HD 243980  & $\zeta$ Per  & BI1b\\
 HD 214993  & 12 Lac  &  B1.5III\\
 HD 148605  & 22 Sco   &  B2V\\
 HD 886  & $\gamma$ Peg   &  B2IV\\
 HD 29248  & $\nu$ Eri  &  B2III\\
 HD 207330  & $\pi^{2}$ Cyg  &  B2.5III\\
 HD 198478  & 55 Cyg  &  B2.5Ia\\
 HD 20365  & 29 Per  &  B3V\\
 HD 160762  & $\iota$ Her  &  B3IV\\
 HD 219688  & $\psi^{2}$ Aqr   &  B5V\\
 HD 147394  & $\tau$ Her   &  B5IV\\
 HD 184930  & $\iota$ Aql  &  B5III\\
 HD 23338  & 19 Tau  &  B6V$^*$\\
 HD 23302  & 17 Tau  &  B6III$^*$\\
 HD 23288  & 16 Tau  &  B7IV\\
 HD 23630  & $\eta$ Tau   &  B7III\\
 HD 214923  & $\zeta$ Peg   &  B8V$^*$\\
 HD 23850  & 27 Tau  &  B8III\\
 HD 196867  & $\alpha$ Del   &  B9V$^*$\\
 HD 176437  & $\gamma$ Lyr   &  B9III\\
 HD 222661  & $\omega^{2}$ Aqr   &  B9.5V\\
 HD 186882  & $\delta$ Cyg   &  B9.5III\\
\end{tabular}
\end{table}

\section{Spectral Classification}

\subsection{General Methodology}

The observed spectrum of Be stars is a composite of
the photospheric absorption spectrum and the spectrum produced by the 
envelope, i.e., an additional continuum component on which 
emission and absorption lines can be superimposed. For most Be stars, 
the contribution of the envelope to the  continuum in the classical 
``classification region''($\lambda \lambda$ 3900\,--\,4900 \AA) is not 
very important (Dachs et al. 1989). However, the envelope can still 
contribute emission in the lines of \ion{H}{i}, \ion{He}{i} and
several metallic ions. Weak emission can result in the ``in-filling''
of photospheric lines, while stronger emission results in the appearance
of emission lines well above the continuum level. With sufficient
spectral resolution and a high enough $v \sin i$ the emission lines
appear double peaked.  A 
well-developed ``shell'' spectrum, with a large number of metallic 
absorption lines can completely veil the photospheric absorption spectrum (see
the spectrum of BD +02$^{\circ}$3815 in Fig 3).

Since the relative strengths of several \ion{He}{i} lines intervene
in most classification criteria for the MK system in the spectral range
of interest, the spectral classification of Be stars has always been
considered particularly complicated. In many spectra, the in-filling
of \ion{He}{i} lines affects the main classification criteria. When 
\ion{Fe}{ii} emission is present, several lines which are used as 
classification criteria can be veiled (such as the \ion{Si}{ii} $\lambda
\lambda$ 4128\,--\,4130 \AA\ doublet).

The high resolutions obtainable at high signal-to-noise ratio 
with modern CCD cameras improve 
the situation, since they allow us to disentangle lines that were blended
at the resolutions formerly used for spectral classification. On the
other hand, the improved resolution means that the traditional criteria 
are not always applicable. Our spectra have a much higher resolution
than the 63 \AA\ mm$^{-1}$ plates used by Walborn (1971) to define the
grid for early-type B stars. Given that the 
only acceptable methodological procedure in the MK scheme is the 
comparison of spectra (Morgan \& Keenan 1973), all this results
in a strong dependence on the choice of standard stars. Unfortunately,
the standard stars available for observation are limited by the position
of the observatory and time of the year. Our standard stars, taken from
the lists of Walborn (1973) and Jaschek \& G\'{o}mez (1998) are listed 
in Table 3. We point out
that Jaschek \& G\'{o}mez (1998) give HD 23338 (19 Tau) as a B6V standard 
and HD 196867 ($\alpha$ Del) as B9V, while Morgan \& Keenan (1973) give
them as B6IV and B9IV respectively. At our resolution, neither of the two
spectra can be justifiably classified as main sequence objects and we
endorse the subgiant classification. Indeed HD 23338 looks remarkably 
similar to HD23302 (17 Tau), which is given by Jaschek \& G\'{o}mez (1998)
and Lesh (1968) as the primary B6III standard. 

\def\epsfsize#1#2{0.9#1}
\begin{figure*}
\setlength{\unitlength}{1.0in}
\centering
\begin{picture}(6.0,5.0)(0,0)
\put(-0.8,-0.8){\epsfbox[0 0 2 2]{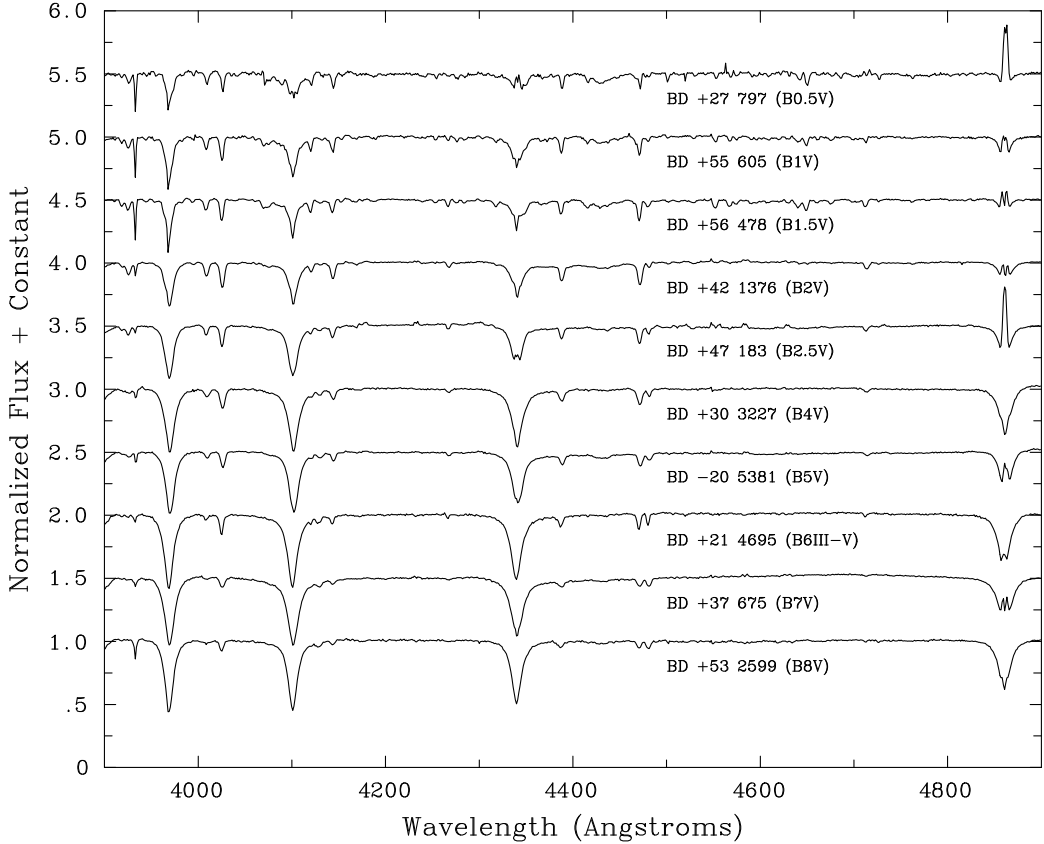}}
\end{picture}
\caption{Spectral Sequence for main sequence stars. 
Note the progressive decline of the He\,{\sc i} 
spectrum from a maximum at B1-B2 and the increase of Mg\,{\sc ii}
$\lambda$4481\AA\ with spectral type. We note that the spectrum of BD +37
675 shows Mg\,{\sc ii} $\lambda$4481\AA $\simeq$ He\,{\sc i}
$\lambda$4471\AA, which defines B8. However, since He\,{\sc i} 
$\lambda \lambda$4711, 4009, 4121 
\AA\ and C\,{\sc ii} $\lambda$4267 \AA\ are still visible, we
have preferred a B7V spectral type, assuming that He\,{\sc i} 
$\lambda$4471
\AA\ is partially filled-in.  If this is not the case,
an intermediate spectral type (B7.5V) would seem necessary}
\end{figure*}

\def\epsfsize#1#2{0.9#1}
\begin{figure*}
\setlength{\unitlength}{1.0in}
\centering
\begin{picture}(6.0,3.0)(0,0)
\put(-0.8,-0.8){\epsfbox[0 0 2 2]{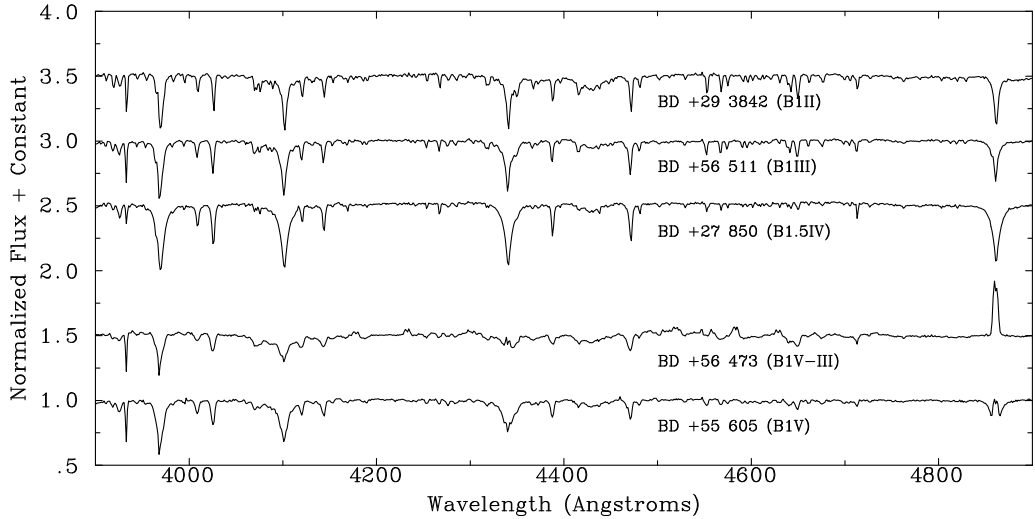}}
\end{picture}
\caption{Luminosity Sequence near B1.  Note the 
increase in the metallic (mainly
\ion{O}{ii}) spectrum with luminosity class, not so obvious in BD +27 850
because of the later spectral type. The emission veiling in BD +56 473 is
too strong to allow an exact classification, even though the strong
\ion{O}{ii} $+$ \ion{C}{iii} near $\lambda$4650\AA\ seems to favour the
giant classification}
\end{figure*}

For classification purposes, we compared the spectra of the Be stars
with those of the standard stars in the interval $\lambda\lambda$ 
3940\,--\,4750
\AA\ both at full instrumental resolution and binned to 1.2 \AA/pixel
to mimic the resolution of photographic plates.  The comparison was
done both ``by eye'' and using the measured equivalent widths of 
the relevant features.
All the spectra have
been classified in this scheme without previous knowledge of spectral
classifications existent in the literature.  The derived spectral 
types are listed in Table 4.  Representative spectral and luminosity
sequences are shown in Figs. 1 and 2, and some peculiar spectra from
the sample are
shown in Fig 3.
We find that the 
accuracy that can be obtained in the classification depends in part 
on the spectral type of the object, as described below.

\def\epsfsize#1#2{0.9#1}
\begin{figure*}
\setlength{\unitlength}{1.0in}
\centering
\begin{picture}(6.0,3.0)(0,0)
\put(-0.8,-0.8){\epsfbox[0 0 2 2]{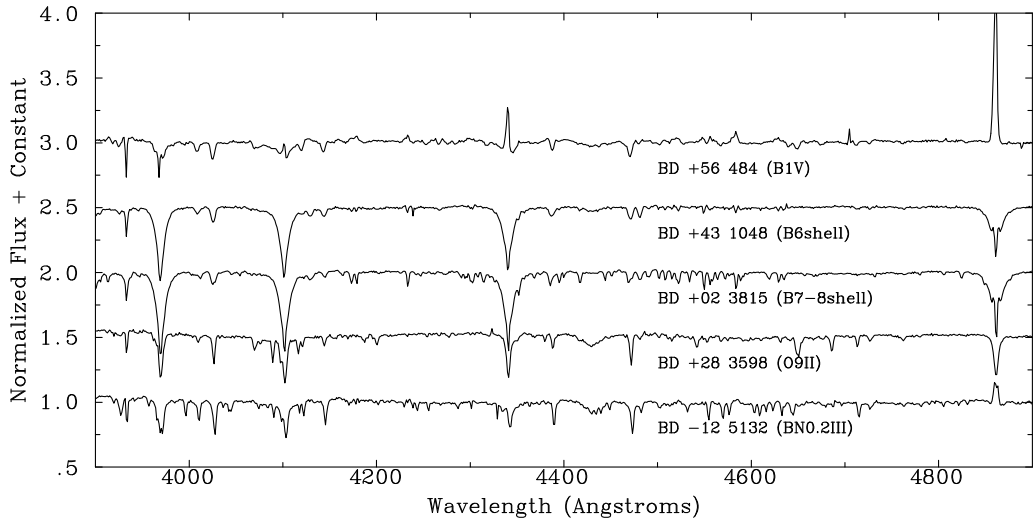}}
\end{picture}
\caption{
Some peculiar spectra in our sample. BD +56$^{\circ}$484 
(B1Ve) 
has a very strong emission spectrum, with abundant \ion{Fe}{ii} 
features.  
For late
types, \ion{Fe}{ii} is only seen in shell spectra, like that of 
BD +43$^{\circ}$1048 (B6IIIshell). BD +02$^{\circ}$3815 (B7-8shell) 
shows 
a fully developed shell spectrum. The earliest spectrum in our sample is
that of BD +28$^{\circ}$3598 (O9II), where no evidence of emission is
detectable. Finally, BD $-$12$^{\circ}$5132 (BN0.2IIIe) is N-enhanced, 
showing strong \ion{N}{ii} $\lambda\lambda$ 3995, 4044, 4242, 4631 \AA\
among others, while \ion{C}{ii} $\lambda\lambda$ 4076, 4650 \AA\ are 
absent.}
\end{figure*}

\subsection{Early B stars}

For stars earlier than B3, the classification can be adequately performed
using as main indicators the \ion{Si}{iv} and \ion{Si}{iii} lines. The 
strength of these lines and of the \ion{O}{ii} spectrum is very sensitive 
to temperature and luminosity variations. Moreover, the emission spectrum
does not generally extend shortwards of $\sim \lambda 4200$ \AA. As a 
consequence, most of our determinations in this spectral range are very
secure. Even though the spectral grid is finer than at later types, we 
believe that most objects are correctly classified to the sub-subtype, i.e.,
a spectrum classified as B0.7III is certainly within the range 
B0.5III\,--\,B1III and both B0.5III and B1III look inadequate 
classifications. The luminosity classification is also secure in this 
range, where we are able to differentiate clearly between giants and
main sequence objects.

\subsection{Late B stars}

For stars later than B5, all the classification criteria available are
strongly affected by the presence of an emission continuum. This has 
resulted in our determinations for this spectral range being slightly
less secure than for earlier spectral types. However, with few exceptions,
we have been able to assign a spectral type to the correct subtype.
This means that we feel that a star classified as B6V would be inadequately
classified as B5 or B7. The luminosity classification is slightly less 
certain. For this reason, we have resorted to using two extra criteria, 
namely, the number of Balmer lines that could be resolved in the spectrum
as it approaches the Balmer discontinuity and the full width
half maximum of H$\theta$ (3797 {\AA}), 
which, among the standard stars, correlates strongly with luminosity class 
at a given spectral type and is not generally affected by emission
in the Be stars. Overall, the three methods do not show strong
discrepancies and our luminosity classification can be considered 
secure, at least to the point of discriminating between giant and
main sequence stars.

\def\epsfsize#1#2{0.9#1}
\begin{figure*}
\setlength{\unitlength}{1.0in}
\centering
\begin{picture}(6.0,3.0)(0,0)
\put(-0.8,-0.8){\epsfbox[0 0 2 2]{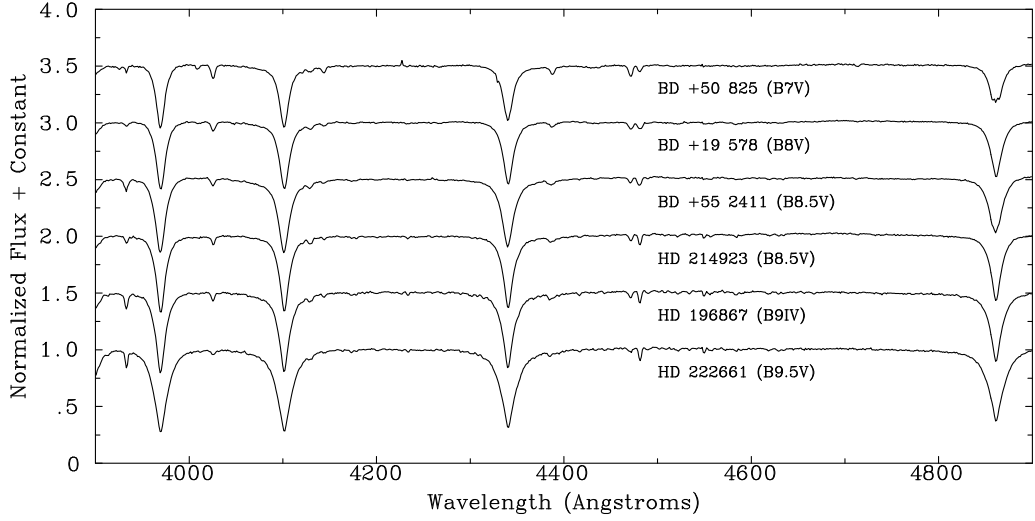}}
\end{picture}
\caption{
Spectral sequence for late B-type stars. In BD +50$^{\circ}$825
(B7Ve), \ion{He}{i} $\lambda$4471 \AA\ is still stronger than 
\ion{Mg}{ii} $\lambda$4481 \AA, but in  BD +19$^{\circ}$578, the two lines
have about the same strength, making the object B8V. \ion{He}{i} $\lambda
\lambda$ 4009, 4121 \AA\ are not visible in this spectrum. Both 
BD +55$^{\circ}$2411 and HD 214923 have \ion{Mg}{ii} $\lambda$4481 {\AA}
stronger than  \ion{He}{i} $\lambda$4471 \AA, which makes them later, even
though HD 214923 is given as B8V standard. Comparison of HD 214923 with 
HD 196867 (B9IV) and HD 222661 (B9.5V) shows that it is not later
than B9V. BD +55$^{\circ}$2411 is earlier than HD 214923, since it 
shows stronger \ion{He}{i} $\lambda \lambda$ 4026, 4387 \AA\ and no
sign of \ion{Fe}{ii} absorption. Therefore a spectral type B8.5V 
seems justified. This spectrum shows no sign of emission over the
whole classification range.
}
\end{figure*}

Lesh (1968) defines B8V by the condition \ion{Mg}{ii} $\lambda$4481 
\AA\ $\simeq$ \ion{He}{i} $\lambda$4471 \AA. However, as can 
be seen in Fig. 4, the spectrum of HD 214923 ($\zeta$ Peg), given by 
Jaschek \& G\'{o}mez (1998) as B8V standard, shows the \ion{Mg}{ii} line 
to be clearly stronger than the \ion{He}{i} line. 
Therefore, this object must be of a later spectral type. Comparison
with HD 196867 ($\alpha$ Del) shows
that this object is not later than B9V. We have taken the spectral
type of this object to be B9V, though we believe that B8.5V could be
an adequate interpolation. No object in our sample is so late as
HD 214923 and therefore we assign to BD +55$^{\circ}$2411, the only
object in the sample with \ion{Mg}{ii} $\lambda$4481 \AA\ clearly 
stronger than \ion{He}{i} $\lambda$4471 \AA\, a spectral type B8.5V.

In the spectral region B7-B8, where the ratio between \ion{Mg}{ii} 
$\lambda$4481 \AA\ and \ion{He}{i} $\lambda$4471 \AA\ is the only
classification criteria, in-filling can strongly affect the derived
spectra. For that reason, we have also used the strength of the 
\ion{Si}{ii} doublet and of the whole \ion{He}{i} spectrum as 
additional information. Moreover, at this resolution, \ion{C}{ii}
$\lambda$4267 \AA\ is visible in main sequence stars up to spectral
type B7V. \ion{Fe}{ii} $\lambda$4232 \AA\ starts to be visible in
the B8V spectra, but we have not used it as a classification indicator
since it can also be a weak shell line.

\subsection{Shell stars}

For the purpose of spectral classification, we have only marked as
``shell'' stars those showing narrow absorption Fe {\sc ii} 
lines, either on top of
emission lines or blanketing the continuum. Several other stars show
double-peaked emission split by an absorption core in some Balmer lines,
but this emission is still inside the photospheric absorption feature and
does not reach the continuum (e.g., BD +37 675 and BD +42 1376 in Fig. 1). 
The shell definition is not applicable since no iron lines 
are visible (see Hanuschik 1995). 
Exceptions could be BD +47
857 and BD +50 3430 
which seem to show absorption cores in some Fe {\sc ii} 
emission lines and therefore
could be shell stars.  We will revisit this question in future papers when
we discuss the emission line spectra of the objects.

\begin{table}
\caption {Measured Spectral Type and $v \sin i$ for the Be star sample}
\begin{tabular}{lll}
OBJECT & Spec. type & $v \sin i$\\ 
\hline 
CD -28 14778 & B2III & $153 \pm 21$\\ 
CD -27 11872 & B0.5V-III & $224 \pm 33$\\ 
CD -27 16010 & B8IV & $187 \pm 32$\\ 
CD -25 12642 & B0.7III & $77 \pm 18$\\ 
CD -22 13183 & B7V & $174 \pm 10$\\ 
BD -20 05381 & B5V & $202 \pm 10$\\ 
BD -19 05036 & B4III & $121 \pm 10$\\ 
BD -12 05132 & BN0.2III & $120 \pm 43$\\ 
BD -02 05328 & B7V & $151 \pm 15$\\ 
BD -01 03834 & B2IV & $168 \pm 34$\\ 
BD -00 03543 & B7V & $271 \pm 54$\\ 
BD +02 03815 & B7-8shell & $224 \pm 14$\\ 
BD +05 03704 & B2.5V & $221 \pm 10$\\ 
BD +17 04087 & B6III-V & $156 \pm 39$\\ 
BD +19 00578 & B8V & $240 \pm 70$\\ 
BD +20 04449 & B0III & $81 \pm 11$\\ 
BD +21 04695 & B6III-V & $146 \pm 10$\\ 
BD +23 01148 & B2III & $101 \pm 10$\\ 
BD +25 04083 & B0.7III-B1II & $79 \pm 11$\\ 
BD +27 00797 & B0.5V & $148 \pm 74$\\ 
BD +27 00850 & B1.5IV & $112 \pm 25$\\ 
BD +27 03411 & B8V & $194 \pm 10$\\ 
BD +28 03598 & O9II & $90 \pm 12$\\ 
BD +29 03842 & B1II & $91 \pm 16$\\ 
BD +29 04453 & B1.5V & $317 \pm 20$\\ 
BD +30 03227 & B4V & $218 \pm 21$\\ 
BD +31 04018 & B1.5V & $211 \pm 11$\\ 
BD +36 03946 & B1V & $186 \pm 21$\\ 
BD +37 00675 & B7V & $207 \pm 29$\\ 
BD +37 03856 & B0.5V & $104 \pm 17$\\ 
BD +40 01213 & B2.5IV & $128 \pm 20$\\ 
BD +42 01376 & B2V & $196 \pm 10$\\ 
BD +42 04538 & B2.5V & $282 \pm 10$\\ 
BD +43 01048 & B6IIIshell & $220 \pm 20$\\ 
BD +45 00933 & B1.5V & $148 \pm 16$\\ 
BD +45 03879 & B1.5V & $193 \pm 10$\\ 
BD +46 00275 & B5III & $113 \pm 21$\\ 
BD +47 00183 & B2.5V & $173 \pm 12$\\ 
BD +47 00857 & B4V-IV & $212 \pm 16$\\ 
BD +47 00939 & B2.5V & $163 \pm 12$\\ 
BD +47 03985 & B1-2shell & $284 \pm 20$\\ 
BD +49 00614 & B5III & $90 \pm 27$\\ 
BD +50 00825 & B7V & $187 \pm 10$\\ 
BD +50 03430 & B8V & $230 \pm 15$\\ 
BD +51 03091 & B7III & $106 \pm 10$\\ 
BD +53 02599 & B8V & $191 \pm 23$\\ 
BD +55 00552 & B4V & $292 \pm 17$\\ 
BD +55 00605 & B1V & $126 \pm 35$\\ 
BD +55 02411 & B8.5V & $159 \pm 90$\\ 
BD +56 00473 & B1V-III & $238 \pm 19$\\ 
BD +56 00478 & B1.5V & $157 \pm 12$\\ 
BD +56 00484 & B1V & $173 \pm 16$\\ 
BD +56 00493 & B1V-IV & $270 \pm 10$\\ 
BD +56 00511 & B1III & $99 \pm 14$\\ 
BD +56 00573 & B1.5V & $250 \pm 58$\\ 
BD +57 00681 & B0.5V & $147 \pm 49$\\ 
BD +58 00554 & B7V & $229 \pm 10$\\ 
BD +58 02320 & B2V & $243 \pm 20$\\
\end{tabular}
\end{table}

\subsection{Distribution of Spectral Types within the Sample}

In Fig. 5 we plot histograms showing the total number of
objects of each spectral type and luminosity class in our sample.
For the purposes of the histograms and the discussion 
that follows
we have classified the two early objects we give luminosity class II
as III, and the objects we classify as ``III-V'' as IV.
Apart from the very earliest objects, which are all giants, the distributions
of objects between the three luminosity classes with respect to spectral type
are similar.  This is probably a selection effect in that there are
very few early-type stars, so most are far away and our magnitude limited
sample only selects the more luminous giants.  From Fig. 6 we note that these
very early objects are near the magnitude limit of our sample, thus
supporting this interpretation.

\def\epsfsize#1#2{0.9#1}
\begin{figure*}
\setlength{\unitlength}{1.0in}
\centering
\begin{picture}(6.0,4.5)(0,0)
\put(-0.6,-0.5){\epsfbox[0 0 2 2]{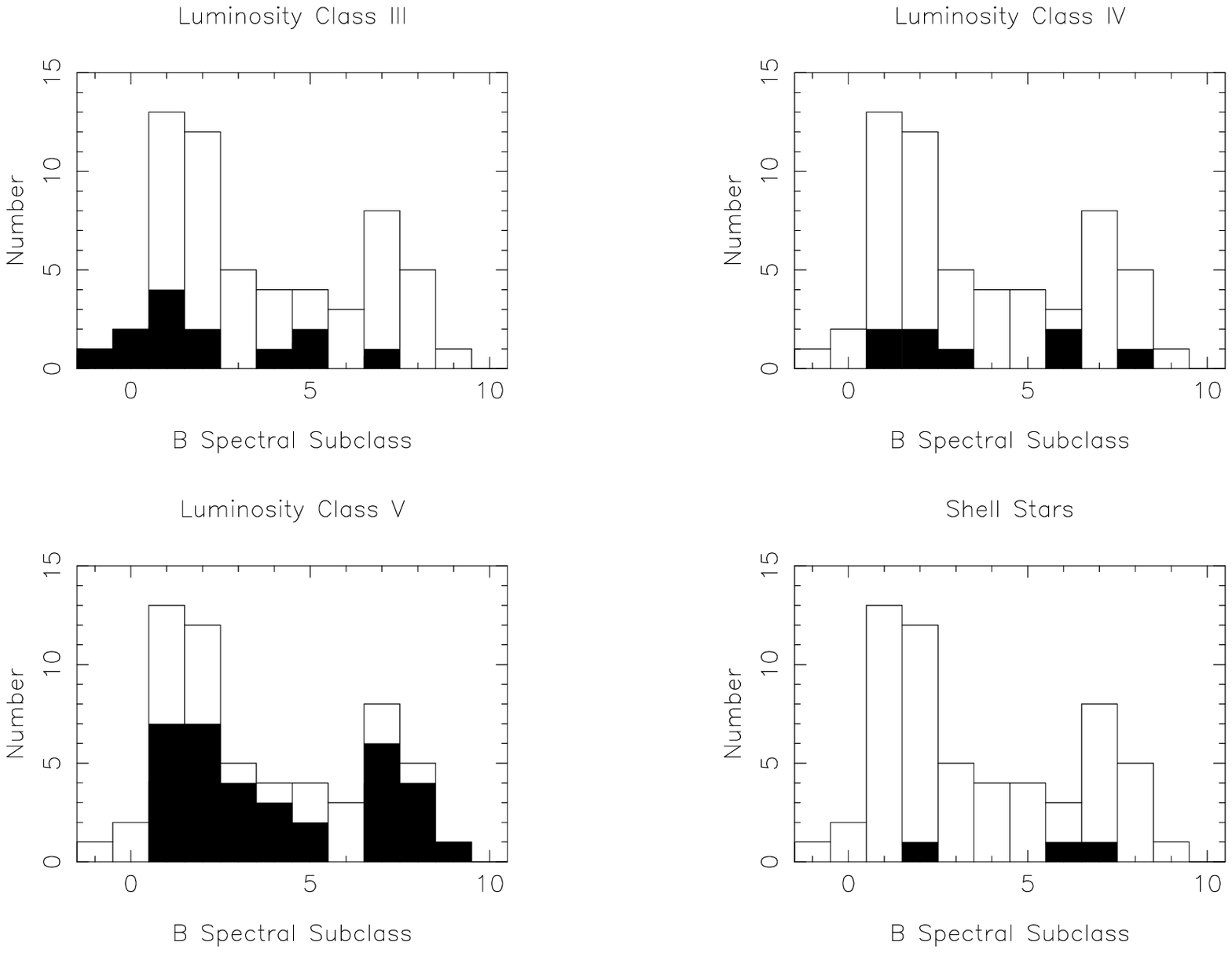}}
\end{picture}
\caption{The distribution of our sample by spectral subclass for
luminosity classes III,IV and V and the shell stars (solid area).
Also plotted on each histogram (hollow area) is the overall distribution
neglecting luminosity class.  The data have been binned into bins n
containing objects in the range range B(n-1).5 to B(n).4.  
Therefore a B0.2 object will appear in the B0 bin, however a B0.5 object
appears in the B1 bin.}
\end{figure*}

Considering the sample as a whole it is interesting to note 
the majority (34 out of 58)
objects are dwarfs, with only 13 out of 58 unambiguously classified as
giants.  Recall that in Section 2 our selection criteria were
designed to give an equal number of these objects.  This implies
that in spectral types given by Jaschek \& Egret (\cite{je82})
many objects that are classified as luminosity class III should 
in fact be classified as IV or V.  In this case it is unlikely that this
is a bias created by our selection of objects from the catalogue,
as our magnitude limit would be expected to select
the more luminous giants over the dwarfs.

\def\epsfsize#1#2{0.8#1}
\begin{figure}
\setlength{\unitlength}{1.0in}
\centering
\begin{picture}(3.0,3.2)(0,0)
\put(-0.6,-0.1){\epsfbox[0 0 2 2]{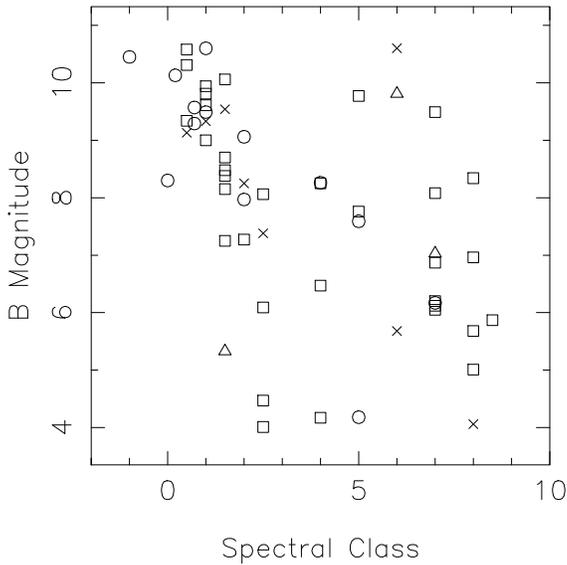}}
\end{picture}
\caption{Apparent B magnitude versus B spectral subclass for the sample.
Luminosity class is indicated by the symbols, where a circle indicates III,
a cross IV and a square V.  Objects with a shell spectrum are indicated by
the triangles.}
\end{figure}

\section{Rotational Velocities}

\subsection{Methodology}

In order to derive rotational velocities we attempted to fit Gaussian 
profiles to
4 He{\sc I} lines at 4026 {\AA},
4143 {\AA}, 4387 {\AA} and 4471 {\AA} in the spectra.  This was
successful for most of the objects, 
although for a small number only 2 or 3 lines could be reliably fitted due to
contamination by nearby emission or absorption features.
The profile full widths at half maximum were
converted to $v \sin i$ using a fit to the 4471 {\AA} full width half 
maximum - $v \sin i$ correlation of Slettebak et al. (\cite{s75}).
Making the appropriate correction for the differing central wavelengths of
each line, the fits employed were:
\begin{equation}
v \sin i = 41.25 F(4471) {\rm km/s}
\end{equation} 
\begin{equation}
v \sin i = 42.03 F(4387) {\rm km/s}
\end{equation}
\begin{equation}
v \sin i = 44.51 F(4143) {\rm km/s}
\end{equation}
\begin{equation}
v \sin i = 45.82 F(4026) {\rm km/s}
\end{equation}
where $F(\lambda)$ is the full width half maximum in {\AA} at a wavelength
of $\lambda$ {\AA}.  The $v \sin i$ quoted in Table 4 is the 
mean of those derived from all of the fitted lines for each object 
after correction for
the mean instrumental velocity dispersion of 55 km/s (determined from
measurements of interstellar lines in the spectra).  The errors reflect the
dispersion in the measured full width half maxima, with the minimum
error set at 10 km/s.

\def\epsfsize#1#2{0.57#1}
\begin{figure}
\setlength{\unitlength}{1.0in}
\centering
\begin{picture}(3.0,3.0)(0,0)
\put(-0.6,-0.5){\epsfbox[0 0 2 2]{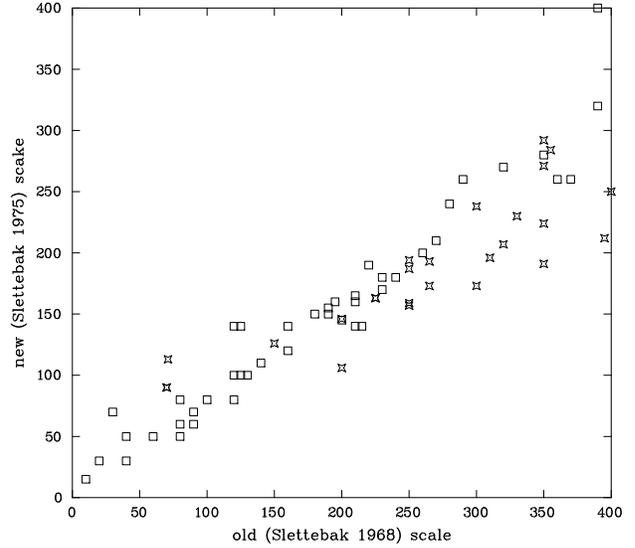}}
\end{picture}
\caption{Comparison of ``old'' (Slettebak 1968) and ``new'' (Slettebak 1975)
$v \sin i$ scales for B stars (square symbols) and our measurements of Be stars
(new scale) versus Bernacca \& Perinottot's (1970,1971) (old scale)
data for the same objects (star symbols).}
\end{figure}

26 of the objects in our sample have previously measured $v \sin i$'s in the
compilations of Bernacca \& Perinotto (\cite{bp70},\cite{bp71}).  A comparison
of the historical $v \sin i$'s with those we derive shows the historical
values typically $\sim 20$ per-cent greater than our values.  However
Bernacca \& Perinotto (\cite{bp74}) state that their $v \sin i$'s 
are referenced
to the scale of Slettebak (\cite{s68}) whereas our measurements are
instead referenced to Slettebak et al. (\cite{s75}).  Figure 7 of that paper
shows that the new (1975) scale derives $v \sin i$'s some 15-20 per-cent
smaller for B stars than the old (1968) scale and so the discrepancy may
simply be understood to be caused by our use of a more modern
$v \sin i$ calibration.  To make this clear Figure 7 re-plots Figure 7
of Slettebak et al. (\cite{s75}) with their standard stars marked as
squares and our sample marked as crosses.  No significant difference
is apparent between the two distributions.

\subsection{Distribution of Rotational Velocities within the sample}

\def\epsfsize#1#2{0.57#1}
\begin{figure}
\setlength{\unitlength}{1.0in}
\centering
\begin{picture}(3.0,6.1)(0,0)
\put(-0.0,-0.0){\epsfbox[0 0 2 2]{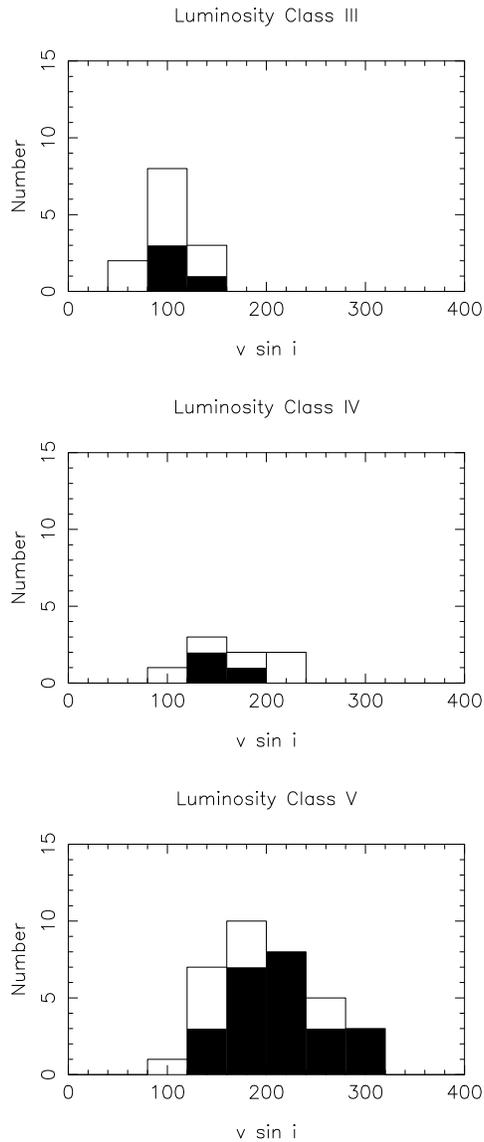}}
\end{picture}
\caption{$v \sin i$ distributions for all objects in the sample (hollow
plus filled areas) and objects in a volume limited subset of the
sample (filled area only)}
\end{figure}

In Figure 8 (hollow plus filled areas) 
we plot the distribution of $v \sin i$ versus spectral
type within the sample for luminosity classes III,IV and V.  
It is important to identify whether there are are biases
in the $v \sin i$ values within the sample. There are two effects 
that may lead to a relationship between $v \sin i$ and brightness
for Be stars and in a flux limited sample
we would of course expect an 
inherent bias towards intrinsically brighter objects.

The first effect that would lead to a bias towards rapidly rotating
Be stars is described by Zorec \& Briot (\cite{zb97}).  The
more rapidly rotating stars will suffer greater deformation
than the slower rotators. 
Although the bolometric luminosity from an object is
of course conserved, the effect of rotation is to make the spectrum appear
cooler.  Since the peak of B star spectra is in the UV, this will make
the optical flux brighter for most aspect ratios (Collins et al. 1991, 
Porter 1995), leading to the preferential selection of more
rapidly rotating stars in a magnitude limited sample.

In addition it is well known that the emission produced in the circumstellar
envelope of Be stars leads to an increase in their optical flux.
Observations of phase changes from non-Be to Be
typically show increases of 
$0.1-0.2$ magnitudes (Feinstein 1975, Apparao 1991).
This is due to reprocessing of the UV radiation from the underlying
Be star into optical and infrared light by the disk.  Assuming a relation
between disk size and excess optical flux we would therefore
expect stars with larger circumstellar disks to be preferentially
selected by our flux limited sample.  Also assuming that
the sizes of Be star disks are likely to be
sensitive to the stellar rotational velocity, then we would
expect the most rapidly rotating stars to show the strongest optical
excess.  This again would lead to a bias towards rapidly
rotating Be stars in a magnitude limited sample.

\def\epsfsize#1#2{0.8#1}
\begin{figure}
\setlength{\unitlength}{1.0in}
\centering
\begin{picture}(3.0,3.2)(0,0)
\put(-0.6,-0.1){\epsfbox[0 0 2 2]{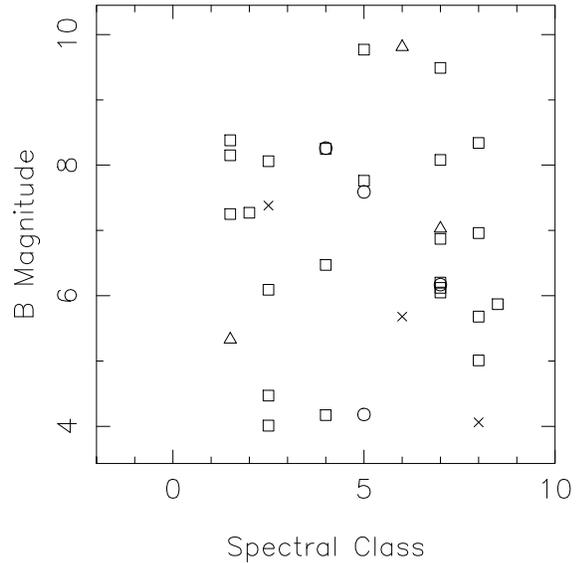}}
\end{picture}
\caption{Apparent B magnitude versus B spectral subclass for the volume
limited sample.
Symbols as per Figure 6}
\end{figure}

In order to test whether our sample is biased in this way
we must compare it to a volume limited subset.  This can be
created from our sample by 
using the absolute magnitudes (Schmidt-Kaler \cite{sch82}) 
derived
from the spectral and luminosity classes to select objects
that lie within the volume defined by the absolute  
magnitude limit for the intrinsically faintest
objects (B9V) at the apparent magnitude limit ($B \sim 11$).
The resulting volume limited subsample contains 34 objects out
of our original 58.  We plot the B magnitude distribution of this
subsample in Fig. 9.  Note how the volume limiting naturally cuts
out the objects with faint apparent magnitudes at early spectral
types  (c.f. Fig. 6).  
The $v \sin i$ distribution of the volume limited subsample
is plotted in Fig. 8 (filled area only) to allow comparison
with the total sample (filled plus hollow area).
In order to
compare the distributions we use a Kolmogorov-Smirnov (KS)
test between the volume-limited and total samples.  This shows that
the probability of the distribution of $v \sin i$ two samples being  
the same is 80\% for luminosity class III, 99\% for luminosity class IV,
and 95\% for luminosity class V.  There is therefore no
statistical evidence for any bias in the $v \sin i$ values for
all three luminosity classes in the sample.  What is clear from
Figure 8 is that there is a considerably 
lower mean $v \sin i$ for luminosity class III as opposed to
class V Be stars.  The astrophysical interpretation of
this result is discussed in Steele (1999).

\section{Conclusions}

We have selected a sample of 58 Be stars of spectral types O9 to
B8.5 and luminosity classes III to V.  We have classified the
spectra by comparison with standards observed at the same
time and derived $v \sin i$ values by measuring the
FWHM of 4 He{\sc i} lines.
Although our initial
criteria were designed to select equal numbers of dwarfs and
giants, in fact our sample is dominated by dwarfs.  The few giants
within the sample tend to have
lower rotational velocities than the dwarfs. By comparison with
a volume limited subsample we have shown that this is not a selection
effect.

The sample defined in this paper has also been observed at
other optical and infrared wavelengths.  The results of those
observations will be presented in future papers, where the
line emission detected 
will be used in an attempt to correlate the properties of
the circumstellar disk with those derived in this paper.

\begin{acknowledgements}

This research has made use of the SIMBAD database of the CDS, Strasbourg.
The Isaac Newton Telescope is operated on the Island of La Palma by the
Isaac Newton Group in the Spanish Observatorio del Roque de los Muchachos
of the Instituto de Astrofisica de Canarias.  IN and JSC were
supported by PPARC.

\end{acknowledgements}

\end{document}